\documentclass[aps,prl,twocolumn,superscriptaddress, show pacs]{revtex4-1}
\usepackage{bm, amsmath}
\usepackage[pdftex]{graphicx}
\usepackage{color}%font colors
\usepackage{epstopdf}
\usepackage{graphicx}% Include figure files
\usepackage{dcolumn}% Align table columns on decimal point
\usepackage{bm}% bold math
\begin{document}

%\title{{Skyrmion racetracks based on the enhanced functionality of non-axisymmetric chiral skyrmions}}

\title{Current-induced shuttlecock-like movement of non-axisymmetric chiral skyrmions}

%\title{\textcolor[rgb]{1,0,0}{Spintronics via non-axisymmetric chiral skyrmions}}

\author{Remi Murooka}

\affiliation{%
Department of Physics, Toho University, 2-2-1 Miyama, Funabashi, Chiba, Japan}

\author{Andrey O. Leonov}
\thanks{leonov@hiroshima-u.ac.jp}
\affiliation{Chirality Research Center, Hiroshima University, Higashi-Hiroshima, Hiroshima 739-8526, Japan}
\affiliation{Department of Chemistry, Faculty of Science, Hiroshima University Kagamiyama, Higashi Hiroshima, Hiroshima 739-8526, Japan}
\affiliation{IFW Dresden, Postfach 270016, D-01171 Dresden, Germany} 

\author{Katsuya Inoue}
\thanks{kxi@hiroshima-u.ac.jp}
\affiliation{Chirality Research Center, Hiroshima University, Higashi-Hiroshima, Hiroshima 739-8526, Japan}
\affiliation{Department of Chemistry, Faculty of Science, Hiroshima University Kagamiyama, Higashi Hiroshima, Hiroshima 739-8526, Japan}

\author{Jun-ichiro Ohe}
\thanks{junichirou.ohe@sci.toho-u.ac.jp}
\affiliation{%
Department of Physics, Toho University, 2-2-1 Miyama, Funabashi, Chiba, Japan}

\date{\today}

\begin{abstract}
{
Current-induced motion of non-axisymmetric skyrmions within angular phases of polar helimagnetis with the easy plane anisotropy is studied by micromagnetic simulations.
Such non-axisymmetric skyrmions consist of a circular core and a crescent-shaped  domain-wall region formed with respect to the tilted surrounding state. % are forced to develop an asymmetric shape in order to match their spin pattern with that of the TFM state
A current-driven motion of non-axisymmetric skyrmions exhibits two distinct time regimes: initially the skyrmions rotate towards the current flow direction and subsequently move along the current with the skyrmionic crescent first.
%A current-driven motion of  non-axisymmetric skyrmions exhibits two distinct time regimes: initially the skyrmions rotate towards the current flow direction and subsequently move along with the skyrmionic crescent first. % or the circular core first. 
%
According to the Thiele equation, the asymmetric distribution of the topological charge and  the dissipative force tensor play an important role for giving the different velocities for the circular and the crescent-shaped constituent parts of the skyrmion what underlies such a shuttlecock-like movement. % and velocity dependence on the .
Moreover, the current-velocity relation depends on the tilt angle of the surrounding angular phase what makes in particular the transverse velocity of skyrmions sensitive to their field-driven configurational transformation. 
%
%The current-velocity relation depends on the tilt angle of the surrounding angular phase: %
%the longitudinal velocity  is not affected by the changing magnetic field whereas 
%Particularly, the transverse velocity is sensitive to the field-driven configurational transformation of skyrmions.  % and is distinct for non-axisymmetric skyrmions with opposite core polarities.
%
We also argue the possibility of magnetic racetrack waveguides based on complex interplay of robust asymmetric skyrmions with multiple twisted edge states. 
% We argue that these {specific} skyrmionic states can be employed in racetrack memory devices as an effective alternative to the common axisymmetric skyrmions 	which occur in magnetically saturated states.  
%
%Three different types of edge states formed at the lateral boundaries of the racetrack impose an orientational confinement of non-axisymmetric skyrmions. 
}
\end{abstract}

% %%% PACS numbers
         
\maketitle

\begin{figure}
\includegraphics[width=0.99\columnwidth]{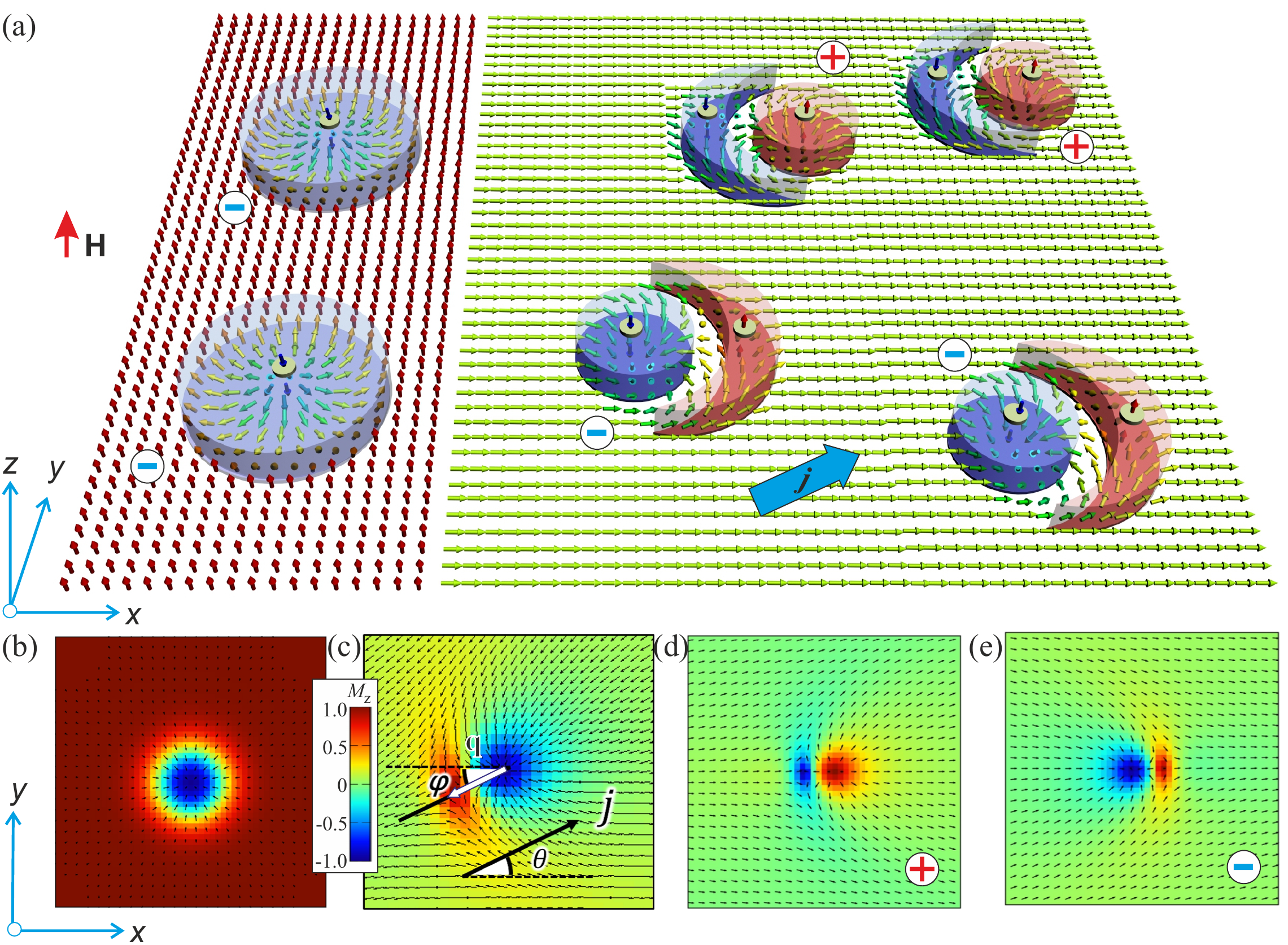}
\caption{ (color online) (a) Schematic of the current-induced motion of  isolated skyrmions in polar magnets.  Magnetic field is applied along $z$-axis and either saturates the surrounding state (left) or due to the competition with the easy-plane anisotropy leads to  the angular phase (right). Skyrmions implanted into these homogeneous states have either axisymmetric (b) or non-axisymmetric (c) shapes. %, correspondingly. 
On the contrary to the saturated state that hosts skyrmions only with the negative polarity (b), the angular phase  accommodates both types of NIS with the magnetization in their cores both along and opposite to the field (d), (e). Note the opposite orientation of the crescent with respect to the cores with opposite polarity.
The color plots indicate $z$-component of the magnetization in AIS (b) and NIS with positive (d) and negative (e) polarities. 
Black arrows are projections of the magnetization on to the $xy$ plane. The spin polarized current comprises angle $\theta$ with $x-$axis while NIS-dipoles $\mathbf{q}$ -- angle $\varphi$ as shown in (c).
\label{art}}
\end{figure}

%general definition
%
\textit{1. Introduction.} Magnetic chiral skyrmions are particle-like topological excitations with complex non-coplanar spin structure \cite{Bogdanov94,Bogdanov89,Nagaosa13,review} recently discovered in bulk non-centrosymmetric helimagnets \cite{Muehlbauer09,Yu10,Wilhelm11,Kezsmarki15,Bordacs2017} and in nanostructures with confined geometries over larger temperature regions\cite{Yu11,Du15,Liang15}. 
Chiral Dzyaloshinskii-Moriya interactions (DMI)\cite{Dz64} protect skyrmions from radial instability \cite{Bogdanov94,Leonov16a} and provide a unique stabilization mechanism to overcome the constraints of the Hobart-Derrick theorem \cite{solitons}. 
Nanometer size of skyrmions, their topological protection and the ease with which they can be manipulated by electric currents \cite{Schulz12,Jonietz10,Hsu17} opened a new active field of research in non-volatile memory and logic devices \cite{Sampaio13,Tomasello14}. % such as racetrack memories.
In particular, in the skyrmion racetrack \cite{Tomasello14,Muller17,Wang16} -- a prominent model for future information technology -- the information flow is encoded in the moving metastable skyrmions \cite{Fert2013}.

The customary  approach to enhance the functionality of the skyrmion-based racetrack memory is  the mechanical patterning of underlying nanosamples.
As an example, suggested devices may feature a regular arrangement of notches to divide the track into a sequence of parking lots for the skyrmions \cite{Fert2013,Fook16}.
An additional nanostrip on top of the racetrack may create an energy barrier along the middle and thus forms two channels for a skyrmion movement \cite{Muller17}. 
This enables the information storage in the lane number of each skyrmion.
Moreover by adding stripes with high magnetic crystalline anisotropy at the edges, one confines the skyrmions inside and prevents their escape from the nanotrack \cite{Lai17}. 
One should also mention elaborate schemes that include, e.g., Y-shaped junctions \cite{Zhang15}.
By these, the skyrmions can be selectively driven into different nanotracks and form complementary data representation.
Moreover, spin logic gates such as the "AND" and "OR" operations based on manipulations of skyrmions can be designed.

An alternative approach to enhance the functionality of the skyrmion-based racetrack memory is to utilize the unique properties of non-axisymmetric isolated skyrmions (NISs) that % as opposed to ordinary 
in particular  may emerge within tilted ferromagnetic phases of polar magnets with the easy-plane anisotropy \cite{Leonov17} (Fig. \ref{art} (a), right).
As compared with the ordinary axisymmetric isolated skyrmions (AISs) within the field-saturated state \cite{Leonov16a} (left side of Fig. \ref{art} (a) and (b)), NIS  may acquire both polarities in their cores (although the vorticity bears the same sign) and thus are subject to the skyrmion Hall effect with opposite shift directions (and hence naturally form two channels, Fig. \ref{art} (d), (e)). %shifted towards opposite directions by the spin-polarized currents (SPC).
%
%subject to the Hall effect with opposite shift directions.
%
Based on the opposite sign of the topological charge, one may call two types of NISs -- skyrmions (Fig. \ref{art} (d)) and anti-skyrmions (Fig. \ref{art} (e)), and consider them as binary data bits for possible practical applications.  % -- similar to skyrmions and anti-skyrmions with the same polarity, but the opposite vorticity. 
Note that skyrmions and antiskyrmions with the same polarity but the opposite vorticity were recently investigated in frustrated magnets with competing exchange interactions \cite{Leonov17b}.
%
%The current-driven dynamics of frustrated skyrmions in a racetrack memory was also shown to be strongly affected by the complex spin states formed at the stripe edges. 

In the present Letter, %
we explore the current-induced dynamics of introduced non-axisymmetric skyrmions. % (a comparative study of dynamycal properties of skyrmions  and anti-skyrmions will be considered elsewhere). 
We show that depending on the direction of the spin-polarized current (SPC) with respect to the skyrmion orientation, %within an angular state,
 NIS undergoes a rotation towards the SPC with its subsequent current-aligned movement. 
The velocity of NIS is effectively regulated by the field magnitude (and thus by the tilt angle of the surrounding angular phase). % and depends on the skyrmion polarity. 
%
%Moreover, NISs develop an attracting skyrmion-skyrmion interaction with the field-dependent inter-skyrmion distance \cite{Leonov17}.
%
We also underline prospects of using NISs in racetrack memory devices.
Anisotropic skyrmion-skyrmion interaction  that depends on their mutual orientation \cite{Leonov17} alongside with the three types of edge states naturally formed at the lateral edges of a racetrack, make NISs effective candidates to be employed in nanoelectronic devices of the next generation in which nanopatterning is boiled down to a minimum. 

\textit{2. Micromagnetic model. }
The equilibrium solutions for NIS are derived  within the standard discrete model of a polar helimagnet  where the total energy is given by:
%We studied minimal-energy states in a stripe of a chiral magnet

\begin{align}
w =  &J\,\sum_{<i,j>} (\mathbf{M}_i \cdot \mathbf{M}_j ) -\sum_{i} \mathbf{H} \cdot \mathbf{M}_i   +K M_z^2 \nonumber\\ 
&- D \, \sum_{i} (\mathbf{M}_i \times \mathbf{M}_{i+\hat{x}} \cdot \hat{y} 
 - \mathbf{M}_i \times \mathbf{M}_{i+\hat{y}} \cdot \hat{x}). 
 \label{model}
\end{align}
$\mathbf{M}_i$ is the unit vector in the direction of the magnetization at the site $i$ of a two-dimensional square lattice and $<i,j>$ denote pairs of nearest-neighbor spins.
$\hat{x}$ and $\hat{y}$ are unit vectors along $x$ and $y$ directions, respectively.
The first term describes the ferromagnetic nearest-neighbor exchange with $J<0$, the second term  is the Zeeman interaction, and the third term is the easy-plane anisotropy with $K>0$.
Throughout the paper, we use the value of $K$ that enables only an angular phase formation, i.e. we omit the regions of the $H-K$ phase diagram that host modulated skyrmion, spiral and elliptical cone phases (see, e.g.,  Fig. 1 in Ref. \onlinecite{Rowland2016} for details).
Dzyaloshinskii-Moriya interaction (DMI) stabilizes NISs with the N\'eel type of the magnetization rotation. 
%
%skyrmions and spirals (cycloids) of N\'eel type with the rotation plane of the magnetization including the wave vector and the polar axis. 
%
The DMI constant $D = J \tan (2\pi/p)$ defines the characteristic size of skyrmions. %period of modulated structures $p$.
In the following simulations, $D$ is set to $0.5J$. %$p$ is set to 14. 
Within model (\ref{model}), AISs exist as metastable excitations of the saturated state for $H > H_{cr} = 2 - K$ (Fig. \ref{art} (b)), while NISs (Fig. \ref{art} (c)) are present for lower fields. % (\textbf{please check}).
We use $K=1.3 D^2/J$, and hence $H_{cr}=0.7D^2/J$. 

The current-driven dynamics of NISs and AISs was simulated using Landau-Lifshitz-Gilbert (LLG) equation:  
%
%\begin{eqnarray}
%\frac{d{\vec{M}}({\bf r},t)}{dt}=-\gamma{\vec{M}}({\bf r},t)\times {\vec{H}}_{\rm eff}+\frac{\alpha}{M_{\rm s}}{\vec{M}}({\bf r},t)\times\frac{%d{\vec{M}}({\bf r},t)}{dt}.\label{LLG}
%\end{eqnarray}
%
%\begin{eqnarray}
%\frac{d{\mathbf{M}_{\mathbf{\mathit{r}}}}}{dt}=-\gamma{\mathbf{M}_{\mathbf{\mathit{r}}}}\times {\mathbf{H}}_{\rm eff}+\frac{\alpha}{M_{\rm s}} % \mathbf{M}_{\mathbf{\mathit{r}}}\times\frac{d{\mathbf{M}_{\mathbf{\mathit{r}}}}}{dt}.
%\label{LLG}
%\end{eqnarray}
%
\begin{eqnarray}
\frac{d{\mathbf{M}_i}}{dt}=-\gamma{\mathbf{M}_i}\times {\mathbf{H}}_{\rm eff}+\frac{\alpha}{M_{\rm s}}  \mathbf{M}_i\times\frac{d{\mathbf{M}_i}}{dt}.
\label{LLG}
\end{eqnarray}
Here $\gamma$ %=1.76\times10^{11}\,$ 1/Ts 
is the gyromagnetic ratio and $\alpha=0.01$ is the Gilbert damping constant.
$H_{\rm{eff}}$ is a local effective magnetic field, which at the site $i$ is given by $\mathbf{H}_{\rm eff}=-\partial w/\partial \mathbf{M}_{\mathbf{\mathit{r}}}$.
The spin transfer torque (STT) consists of the adiabatic part, $\tau=A(\mathbf{j}\cdot\nabla) \mathbf{M}_i$, and the non-adiabatic term, $\tau_{\beta}=A\beta \mathbf{M}_i\times(\mathbf{j}\cdot\nabla) \mathbf{M}_i$, where $A=Jg\mu_B/2eM_s$ is the coefficient proportional to the SPC density. % $j$.
The SPC $\mathbf{j}$ comprises an angle $\theta$ with $x$-axis (Fig. \ref{art} (c)). 
An orientation of NIS is characterized by an angle $\varphi$ between a vector, which connects centers of the circular core and the crescent (skyrmion dipole $\mathbf{q}$),  and $x$-axis (Fig. \ref{art} (c)).

%In order to argue the spin transfer torque, we apply the current in the angle $\theta$ as shown in Fig.\ref{??}.

\begin{figure}
\includegraphics[width=0.95\columnwidth]{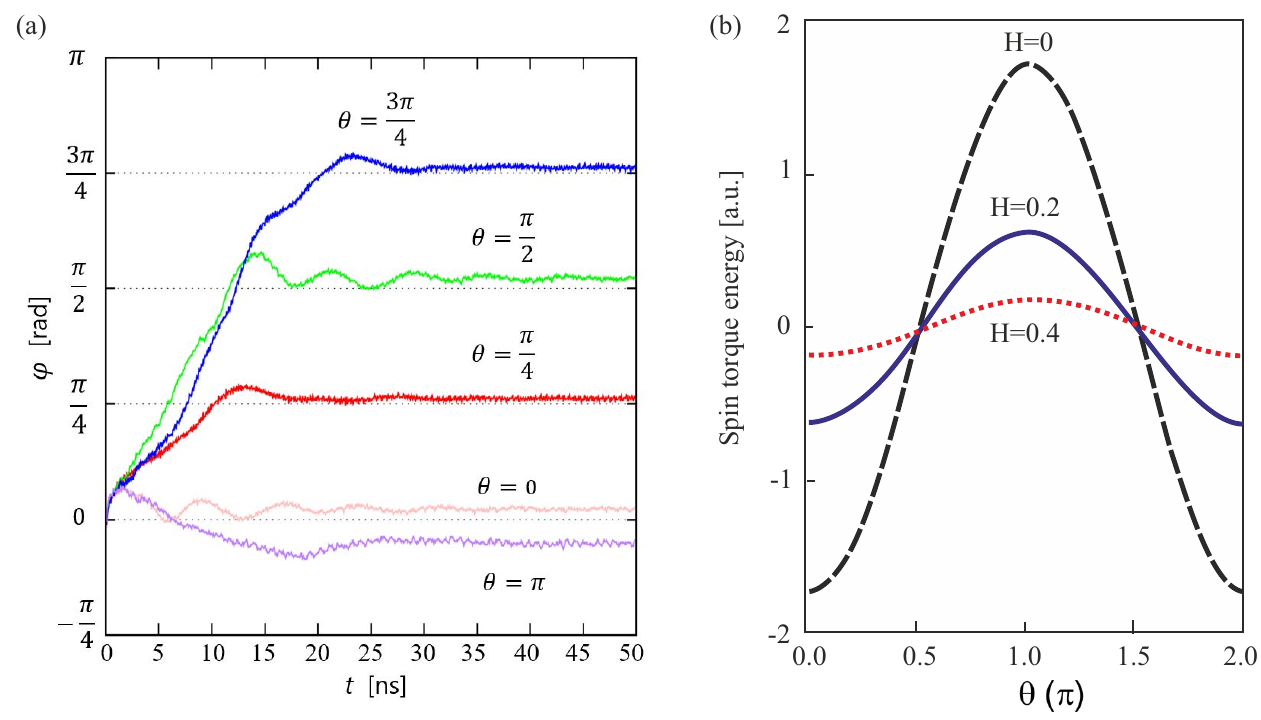}
\caption{ (color online) (a) Current-induced rotation of NISs. Independently on the angle $\theta$, the NIS-dipoles acquire the same orientation angle $\varphi$, i.e. become co-aligned with the SPC direction. Such a forced rotation is explained by the STT energy (b) that is minimized only for the current-aligned movement. 
\label{angle}}
\end{figure}

\textit{3. Current-induced motion  (a shuttlecock-like movement).}
To systematically investigate the current-driven dynamics of NISs, we applied the SPC with different angles with respect to the skyrmion dipole initially oriented with $\varphi=0$ (Fig. \ref{angle} (a)).
It was found that after an initial rotation towards the SPC, NIS-dipoles are always current-aligned with $\varphi=\theta$ (Fig. \ref{angle} (a)).
A relatively small increment of the SPC angle $\theta$ allowed to exclude local minima of the skyrmion orientation with respect to the SPC.
In particular, a NIS motion with its core along the SPC was excluded (although in a numerical experiment of Fig. \ref{angle} (a) such a movement opposite to the SPC appeared to be feasible).  

%Note, that NIS also locally rotate a surrounding angular phase otherwise insensitive to the current (see supplementary video). 

\textit{A. Rotation of NIS towards the SPC direction.}
Figure \ref{angle} (a) shows the time dependence of the dipole angle $\varphi$ for different SPC directions $\theta$. 
As expected, more time is needed to orient NIS along the SPC with increasing angle $\theta$. 
The skyrmion rotation originates from the non-axisymmetric internal structure. 
The spin-torque term $\mathbf{\tau}$ in the LLG equation is represented by
\begin{equation}
\mathbf{\tau}=\mathbf{M}_i\times\left(\mathbf{M}_i\times\left(\mathbf{j}\cdot\nabla\right)\mathbf{M}_i\right)
=\mathbf{M}_i\times\mathbf{H}_{\rm eff}^{\rm STT}.
\end{equation}
and thus the energy gain due to the spin transfer torque $E^{\rm STT}$ can be expressed as $E^{\rm STT} = \int_{\rm sk}d\mathbf{r} \mathbf{M}_i\cdot\mathbf{H}$.
Figure \ref{angle} (b) shows the $E^{\rm STT}$ as a function of $\theta$ featuring minima only along the SPC. %: to minimise the energy, the non-axisymmetric skyrmion rotates by the charge current.
The minima, however, become shallow with the increasing magnetic field and completely disappear for $H>H_{cr}$ with the onset of the field-saturated state that accomodates AISs. 

\begin{figure}
\includegraphics[width=0.99\columnwidth]{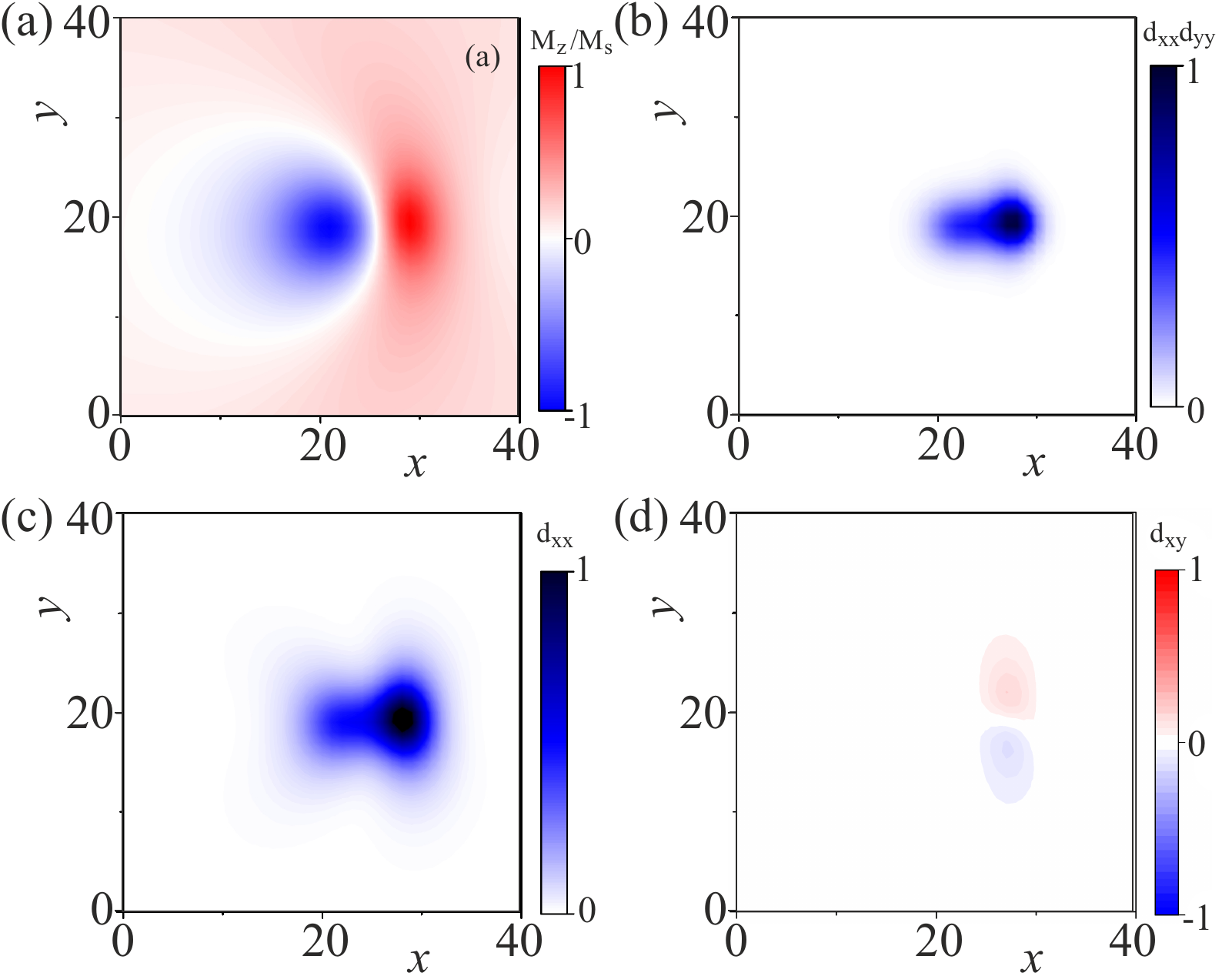}
\caption{ (color online)  Color plots of the $M_z$ component and the dissipative force tensor $d_{xx}d_{yy}$ (b), $d_{xx}$ (c) and $d_{xy}$ (d) as defined by Eq. (\ref{tensor}). The depicted distributions account for a rotational movement of NIS as well as for the translational dynamics in Fig. \ref{velocity} (see text for details). 
\label{thiele2}}
\end{figure}

\begin{figure}
\includegraphics[width=0.99\columnwidth]{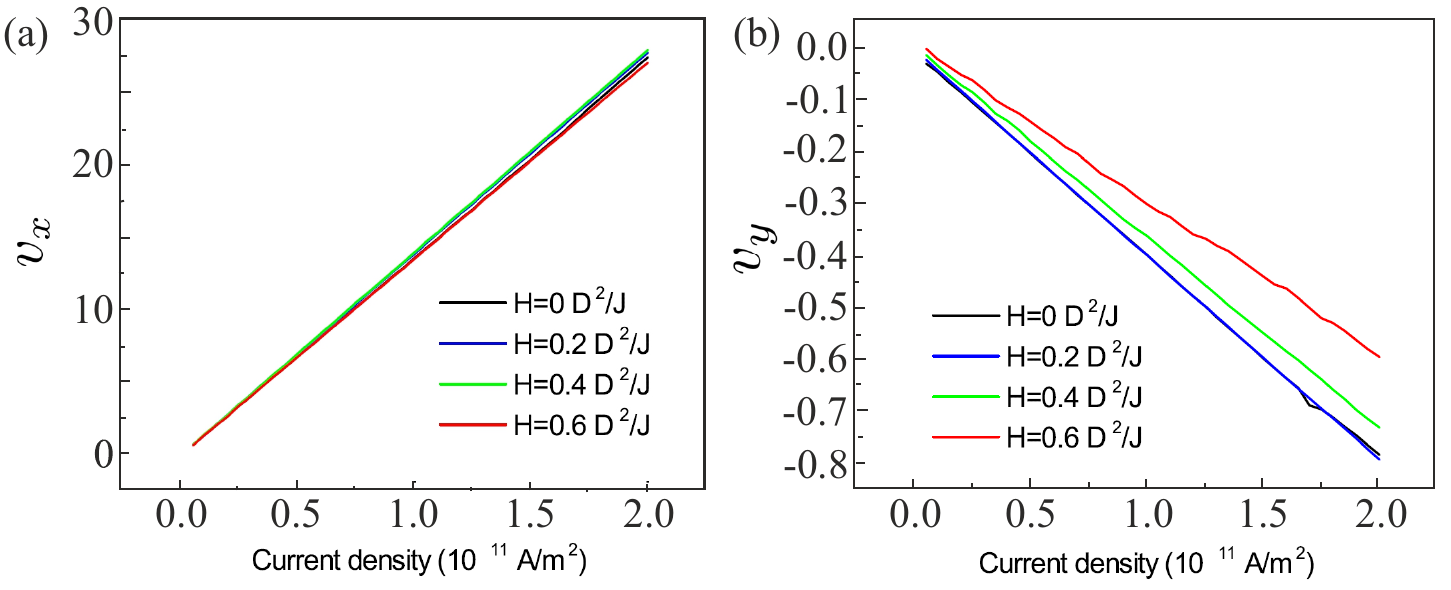}
\caption{ (color online) Longitudinal $v_x$ (a) and transversal $v_y$ (b) velocities of the current-induced motion of NISs for several values of the applied magnetic field and therefore for different angular phases. Whereas $v_x$ is clearly field-independent what complies with the Thiele equation (\ref{thiele}), $v_y$ strongly depends on the skyrmion internal structure (see text for details).
\label{velocity}}
\end{figure}

\textit{B. Translational movement of NISs.}
%
%The velocity of the translational motion along the current direction is proportional to the charge current and the transverse velocity depends on the value of $\beta$ as also observed for AIS \cite{Iwasaki13}.
%
The current-induced translational motion of NIS is well understood in terms of the Thiele equation \cite{Thiele}
%
%From Landau-Lifshitz-Gilbert equation, the Thiele equation is obtained as
%
\begin{eqnarray}
\mathbf{G}\times\left( \mathbf{j}- \mathbf{v} \right) + \mathcal{D}\left( \beta\mathbf{j} - \alpha \mathbf{v} \right)=\nabla V,
\label{thiele}
\end{eqnarray}
where %$\mathbf{j}$ is the current, 
$\mathbf{v}$ is the velocity of the skyrmion and $V$ is the pinning potential.
The gyromagnetic coupling vector $\mathbf{G}=(0,0,G_z)$  equals the topological charge  
\begin{eqnarray}
G_z=\frac{1}{4\pi}\int d\mathbf{r}\frac{1}{M_{\rm s}^3}\mathbf{M}\cdot\left(\partial_x\mathbf{M} \times \partial_y\mathbf{M} \right).
\end{eqnarray}
$\mathcal{D}$ is the dissipative force tensor:
\begin{eqnarray}
\mathcal{D}_{ij}=\frac{1}{2\pi}\int d\mathbf{r}\frac{1}{M_{\rm s}^2}\partial_i\mathbf{M} \cdot \partial_i\mathbf{M} \qquad(i,j=x,y).
\label{tensor}
\end{eqnarray}
which is not symmetric for NISs $(\mathcal{D}_{xx}\neq \mathcal{D}_{yy})$, even
the off-diagonal element has a finite value $(\mathcal{D}_{xy}\neq 0)$.
% for the non-axisymmetric skyrmion.
%
Thus, when the charge current $\mathbf{j}$ is applied, the skyrmion feels both the longitudinal and the Magnus force \cite{Iwasaki13}.

For $\theta=0$,  the longitudinal $(v_x)$ and the transverse $(v_y)$ components of the velocity are represented as 
%
%\begin{equation}
%v_x=\frac{G_z^2+D_{xx}D_{yy}\alpha\beta}{G_z^2+D_{xx}D_{yy}\alpha^2}j,\,\, 
%v_y=\frac{\left( \alpha-\beta \right)D_{xx}G_z}{G_z^2+D_{xx}D_{yy}\alpha^2}j .
%\label{vx}
%\end{equation}
%
\begin{eqnarray}
v_x&=&\frac{G_z^2+D_{xx}D_{yy}\alpha\beta+\xi\beta D_{yx}}{G_z^2+D_{xx}D_{yy}\alpha^2+\xi\alpha D_{xy}}j, \label{vx} \\ 
v_y&=&\frac{\left( \alpha-\beta \right)D_{xx}G_z}{\xi\left\{ G_z^2+D_{xx}D_{yy}\alpha^2 -\xi\alpha D_{xy} \right\} }j .
\label{vy}
\end{eqnarray}
where $\xi=G_z-\alpha D_{xy}$.

We plot the spatial configuration of the dissipative force tensor in Fig.~\ref{thiele2} (b)-(d).
With a sufficiently large value of $d_{xx}d_{yy}=(\partial_x\mathbf{M} \cdot \partial_x\mathbf{M})( \partial_y\mathbf{M} \cdot \partial_y\mathbf{M})$ (plotted in Fig. \ref{thiele2} (b)), 
$v_x=\beta/\alpha$, which is the same as for AISs \cite{Iwasaki13} and is consistent with  the universal $j-v_x$ relation independent of $\beta$ (Fig. \ref{velocity} (a)).
The transverse velocity, on the contrary,  is proportional to $d_{xx}$ (Fig. \ref{thiele2} (c)) and is strongly field-dependent (Fig. \ref{velocity} (b)), since the field affects the spin configuration of NISs. 
One can also see that $d_{xx}d_{yy}$ of the crescent part is larger than that of the circular part (Fig.~\ref{thiele2} (a), (b)).
This asymmetry of $d_{xx}d_{yy}$ underlies the faster velocity of the crescent resulting in the rotational motion (discussed in \textit{3.A} from another point of view). 
We note that the off-diagonal element of the dissipative force tensor $d_{xy}=\partial_x \mathbf{M}\cdot \partial_y\mathbf{M}$ is much smaller than $d_{xx}$ and has opposite sign for upper and lower part of NISs as shown in Fig.~\ref{thiele2} (d).
Therefore, the effect of the off-diagonal element can be neglected.
This is the reason why we dubbed such a motion "a current-induced shuttlecock-like movement". 
The dynamics of the NISs is shown in the Supplemental Material (\cite{SM}). 

\begin{figure*}
\includegraphics[width=1.99\columnwidth]{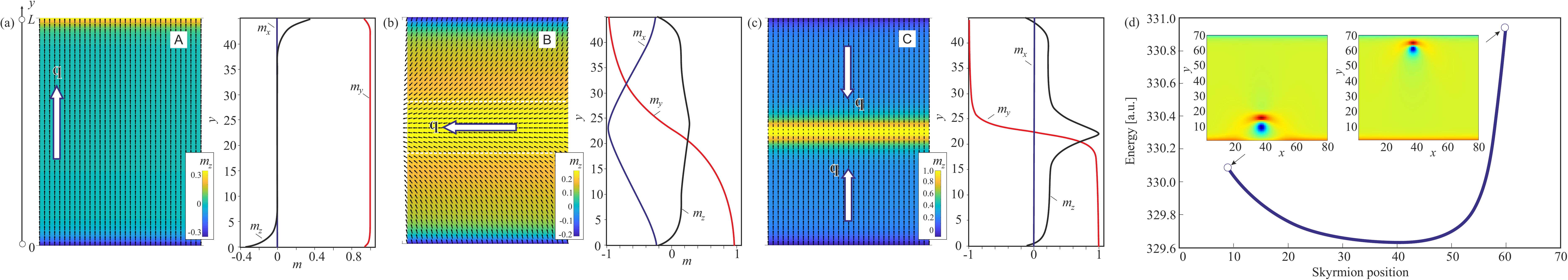}
\caption{ (color online) (a)-(c) Edge states in a nanostripe with the easy-plane uniaxial anisotropy. An applied magnetic field induces consecutive phase transitions A-B-C (see text for details). 
(d) The total energy of the magnetic system (insets) in which the NIS is set on the racetrack. The position of the NIS is defined as the center of the circular vortex part. The total energy becomes larger when the NIS comes near to the edges that assures the propagation of the NISs in the quasi-one dimensional racetrack waveguide. 
\label{edges}}
\end{figure*}

\textit{4. NISs in chiral liquid crystals.} Recently, a low-voltage-driven motion of such non-axisymmetric skyrmions was realized  in chiral nematic fluids \cite{ackerman2015,ackerman2017,sohn2018}  with precise experimental control of both the direction and speed. 
Interestingly, the liquid-crystal (LC) NISs may be realized as a result of the competition between the surface anchoring and an applied electric field: whereas the surface anchoring tends to orient the director $\mathbf{n}(\mathbf{r})$ perpendicular to the confining glass plates, the electric field $\mathbf{E}$ with the negative dielectric anisotropy $\varepsilon_a$ -- parallel to them (see in particular Fig. 1 in Ref. \onlinecite{ackerman2017}).
As a result of such an interplay, one creates an analog of the conical phase around a skyrmion  with the spin angle varying across the thickness. This induces an attractive skyrmion-skyrmion interaction and is experimentally manifested as  chains of NISs \cite{ackerman2015}.
Moreover, the directional motion of LC-NISs is possible as a response to modulated electric fields:
when alternating current with some frequency is applied to a confined skyrmion with the axisymmetric structure (Fig. 1 (b)),
it transforms back and forth into a NIS with the negative polarity (Fig. 1 (c)). NISs with the positive polarity are highly metastable solutions (Fig. 1 (d)) and  have not been realized in the experiment yet. 
Thus, a squirming motion is induced, which can be considered as a LC counterpart of the racetrack memory suggested for magnetic skyrmions.
Note that in the limit of the weak surface anchoring, LC-NISs have precisely the same structure as considered NIS in magnets with the easy-plane anisotropy (Fig. 1). 

We also notice that the results obtained  within the model (\ref{model}) are valid for \textit{bulk} polar magnets with axial symmetry as well as for \textit{thin films} with interface induced DMI.
In particular, bulk polar magnetic semiconductors GaV$_4$S$_8$ \cite{Kezsmarki15} and GaV$_4$Se$_8$ \cite{Bordacs2017} with the
$C_{3v}$ symmetry possess uniaxial anisotropy of easy-axis and easy-plane type, respectively.
Since the magnitude of the effective anisotropy in these lacunar spinels strongly varies with temperature, these material family
provides an ideal arena for the comprehensive study of anisotropic effects on modulated magnetic states \cite{Bordacs2017}.
Skyrmions were also studied experimentally in various systems with interface induced DMI \cite{Romming13,Woo16}.
In these thin film structures, the rotational symmetry can also be broken by different anisotropic environments due to lattice strains
or reconstructions in the magnetic surface layers \cite{Hagemeister16}, as has been discussed recently for the double
atomic layers of Fe on Ir(111) \cite{Hsu16}. These structural anisotropies also promote the formation of asymmetric skyrmions.

\textit{5. Edge states.}
The practical use of NISs in racetracks hinges %crucially depends 
on their interaction with edge states. 
For racetracks with the field-saturated magnetization, the edge states manifest themselves as remnants of the helical spiral and repulse AISs.
For racetracks with the angular phase, however, three different types of edge states can arise (in Fig. \ref{edges} the edge states are marked with capital letters A, B, and C).

\textit{A. Transitions between edge states. }
The type A edge state with collinear in-plane spin components (azimuthal angle of the magnetization is $\psi=\pi/2$) is induced for lower values of the applied magnetic field. 
$z$-component of the magnetization acquires opposite signs at the opposite edges which leads to an asymmetric NIS-edge interaction potential (Fig. \ref{edges} (d)).
With increasing magnetic field, the A-type edge state undergoes the first-order transition into the type B edge state with the rotating magnetization across the racetrack width $L$. 
Once $m_z$ reaches unit value in the racetrack middle, the B-type edge state by the second-order transition  transforms into the type C edge state with $\psi=\pi/2$  (Fig. \ref{edges} (c)).
%
%The former and latter edge states imply a symmetric interaction potential with NIS.

\textit{B. Orientational confinement of NISs.}
Three different types of edge states formed at the lateral boundaries of the racetrack also impose an orientational confinement on non-axisymmetric skyrmions. 
The A-type state implies the perpendicular orientation of a skyrmion dipole $\mathbf{q}$ with respect to the racetrack edges (white arrow in Fig. \ref{edges} (a) and magnetic configurations with NISs located near both stripe edges, Fig. \ref{edges} (d)). At the same time due to the asymmetry of the magnetization distribution within the opposite edges, a NIS will be located closer to one edge than to the other (Fig. \ref{edges} (d)). 
The SPC $j_x$, by inducing the skyrmion rotation, may also initiate a transition A-B between edge states in spite of the B-type state is a metastable solution. 
The B-type state, on the contrary,  may accommodate NISs with $\mathbf{q}||x$. Thus, a moving NIS due to the skyrmion Hall effect will shift the whole stripe with the maximal $M_z$-component (marked by the dashed lines in Fig. \ref{edges} (b)) towards one of the edges. 
The C-type edge state allows two opposite NIS orientations degeneracy of which could be removed by the SPC $j_x$ (Fig. \ref{edges} (c)). 

\textit{6. Conclusions.} 
In conclusion, we examined the current-induced dynamics of non-axisymmetric skyrmions that exist within angular phases of polar helimagnets with the easy plane anisotropy.
We considered a shuttlecock-like movement of NISs that consists in their rotation to coalign with the SPC. 
We succeeded in modifying the current-velocity relation by the field-driven control of the  angle in a surrounding homogeneous state. 
We also speculate that a NIS placed into the racetrack memory with three different types of the edge states not only undergoes an orientational confinement, but can also be used as a current-activated tumbler between edge states.
Our results are not only relevant to the application of magnetic skyrmions in memory technology but also elucidate the fundamental properties of skyrmions and the edge states formed in the angular phases of polar helimagnets \cite{Bordacs2017}.

\section{Acknowledgments}
The authors are grateful to Istvan Kezsmarki and Maxim Mostovoy for useful discussions.
This work was funded by JSPS Core-to-Core Program,
Advanced Research Networks (Japan), JSPS Grant-in-Aid for Research Activity Start-up 17H06889, Grant-in-Aid fort Scientic Research (A) 17H01052 from MEXT, Japan and CREST, JST.
AOL thanks Ulrike Nitzsche for technical assistance.

%\section{Supplementary Movie.}
%

\end{document}